\documentclass[12pt,preprint]{aastex}

\slugcomment{GCA-023}

\shorttitle{Measuring the Chaplygin gas equation of state }
\shortauthors{Alcaniz \& Lima}

\begin{document}


\title{Measuring the Chaplygin gas equation of state from angular and luminosity
distances}


\author{J. S. Alcaniz$^{1,2}$ and J. A. S. Lima$^{1,3}$}

\affil{$^{1}$Departamento de F\'{\i}sica, Universidade Federal do Rio
Grande do Norte, C.P. 1641, Natal - RN, 59072-970, Brasil}

\affil{$^{2}$Observat\'orio Nacional, Rua General Jos\'e Cristino 77, 20921-400, S\~ao
Cristov\~ao, Rio de Janeiro - RJ, Brazil}

\affil{$^{3}$IAG, Universidade de S\~ao Paulo, Rua do Mat\~ao, 1226 -
Cidade Universit\'aria, 05508-900, S\~ao Paulo, SP, Brasil}

\begin{abstract}
An exotic dark component, named generalized Chaplygin gas (Cg) and
parameterized by an equation of state $p = -A/\rho_{Cg}^{\alpha}$
where $A$ and $\alpha$ are arbitrary constants, is one of the
possible candidates for dark energy as well as for a unified
scenario of dark matter/energy. In this paper we investigate
qualitative and quantitative aspects of the angular size -
redshift test in cosmological models driven by such a dark
component. We discuss the prospects for constraining the Cg
equation of state from measurements of the angular size at low and
high redshift radio sources and also from a joint analysis
involving angular size and supernova data. A detailed discussion
about the influence of the Cg on the minimal redshift at which the
angular size of an extragalactic source takes its minimal value is
also presented.
\end{abstract}





\section{Introduction}
The impressive convergence of recent observational facts along
with some apparently successful theoretical predictions seem to
indicate that the simple picture provided by the standard cold
dark matter model (SCDM) is insufficient to describe the present
stage of our Universe. From these results, the most plausible
picture for our world is a spatially flat scenario dominated
basically by cold dark matter and an exotic component endowed with
large negative pressure, usually named dark energy. Despite the
good observational indications for the existence of these two
components, their physical properties constitute a completely open
question at present, which gives rise to the so-called dark matter
and dark energy problems.

Dark matter, whose the leading particle candidates are the axions
and the neutralinos, was originally inferred from galactic
rotation curves which show a general behavior that is
significantly different from the one predicted by Newtonian
mechanics. Dark energy or {\emph{quintessence}}, whose the main
candidates are a cosmological constant $\Lambda$ and a relic
scalar field $\phi$, has been inferred from a combination of
astronomical observations which includes distance measurements of
type Ia supernovae (SNe Ia) indicating that the Universe is
speeding up not slowing down (Perlmutter {\it et al.} 1999; Riess
{\it et al.} 1998; 2004), cosmic microwave background (CMB) data
suggesting $\Omega_{total} \simeq 1$ (de Bernardis et al. 2000;
Spergel et al. 2003) and clustering estimates providing $\Omega_m
\simeq 0.3$ (Calberg et al. 1996; Dekel et al. 1997). While the
combination of these two latter results implies the existence of a
smooth component of energy that contributes with $\simeq 2/3$ of
the critical density, the SNe Ia results require this component to
have a negative pressure thereby leading to a repulsive gravity
(for reviews see Peebles \& Ratra 2003, Padmanabhan 2003).

By assuming the existence of these two dominant forms of energy in
the Universe, one finds that the main distinction between them
comes from their gravitational effects. Cold dark matter
agglomerates at small scales whereas dark energy seems to be a
smooth component. In certain sense, such properties are directly
linked to the equation of state of both components (for a short
review see Lima 2004). On the other hand, the idea of a unified
description for the CDM and dark energy scenarios has received
much attention (Matos \& Ure\~na-Lopez 2000; Davidson, Karasik \&
Lederer 2001; Kasuya 2001; Watterich 2002; Padmanabhan \&
Choudhury 2002). For example, Wetterich (2002) suggested that dark
matter might consist of quintessence lumps while Kasuya (2001)
showed that spintessence-like scenarios are generally unstable to
formation of $Q$ balls which behave as pressureless matter. More
recently, Padmanabhan and Choudhury (2002) investigated such a
possibility via a string theory motivated tachyonic field.

Another interesting attempt of
unification was suggested by Kamenshchik et al. (2001) and developed
by Bili\'c et al. (2002) and Bento et al. (2002).
It refers to an exotic fluid, the so-called Chaplygin gas (Cg), whose equation of state
is given by
\begin{equation}
p_{Cg} = -A/\rho_{Cg}^{\alpha},
\end{equation}
with $\alpha = 1$ and $A$ a positive constant related to the
present day Chaplygin adiabatic sound speed, $v_s^{2} = \alpha
A/\rho_{Cg,o}$, where $\rho_{Cg,o}$ is the current Cg density. In
actual fact, the above equation for $\alpha \neq 1$ constitutes a
generalization of the original Chaplygin gas equation of state
proposed by Bento et al. (2002). The idea of a dark-matter-energy
unification from an equation of state like Eq. (1) comes from the
fact that the Cg or the generalized Chaplygin gas (from now on we
use Cg to denote the Chaplygin gas as well as the generalized
Chaplygin gas) can naturally interpolate between nonrelativistic
matter and negative-pressure dark energy regimes. It can be easily
seen by inserting the Eq. (1) into the energy conservation law
($u_{\mu}T^{{\mu}{\nu}}_{;\nu} = 0$). One finds,
\begin{equation}
\rho_{Cg} = \rho_{Cg_{o}}\left[A_s + (1 - A_s)\left(\frac{R_o}{R}\right)^{3(1 +
\alpha)}\right]^{\frac{1}{1 +
\alpha}},
\end{equation}
where the subscript $o$ denotes present day quantities, $R(t)$ is the cosmological scale
factor, and $A_s = A/\rho_{Cg_{o}}^{1 + \alpha}$ is a quantity related to the present
sound speed for the Chaplygin gas ($v_s^{2} = \alpha A_s$). As can be seen
from the above equations, the Chaplygin gas interpolates between non-relativistic matter
[$\rho_{Cg}(R \rightarrow 0) \propto R^{-3}$] and negative-pressure dark component
regimes [$\rho_{Cg}(R \rightarrow \infty) \propto \rm{const.}$].

Motivated by these potentialities, there has been growing interest
in exploring theoretical (Bordemann \& Hoppe, 1993; Hoppe 1993;
Jackiw 2000; Gonzalez-Diaz 2003a; 2003b; Kremer 2003; Khalatnikov
2003; Balakin et al. 2003; Bilic et al. 2003) and observational
consequences of the Chaplygin gas, not only as a possibility of
unification for dark matter/energy but also as a new candidate for
dark energy only. The viability of such scenarios has been tested
by a number of cosmological tests, including SNe Ia data (Fabris,
Goncalves \& de Souza, 2002; Colistete Jr. et al. 2003; Avelino et
al. 2003; Makler, de Oliveira \& Waga 2003; Bertolami et al.
2004), lensing statistics (Dev, Alcaniz \& Jain 2003; Silva \&
Bertolami 2003; Dev, Alcaniz \& Jain 2004), CMB measurements
(Bento, Bertolami \& Sen 2003a; 2003b; 2003c; Carturan \& Finelli
2002; Amendola et al. 2003), age-redshift test (Alcaniz, Jain \&
Dev 2002), measurements of X-ray luminosity of galaxy clusters
(Cunha, Lima \& Alcaniz 2003), future lensing and SNe Ia
experiments (Avelino {\it et al.} 2003; Silva \& Bertolami 2003;
Sahni et al. 2003), statefinder parameters (Sahni et al. 2003), as
well as by observations of large scale structure (Multamaki,
Manera \& Gaztanaga 2003; Bilic et al. 2003; Be\c{c}a et al. 2003; Bean and Dore 2003;
Avelino et al. 2004). In particular, the latter reference has shown that in the context
of unified dark matter/energy models the onset of the non-linear regime on small
cosmological scales may lead to the breakdown of the background solution, even on large
cosmological scales (if $\alpha$ is not equal to zero).

Although carefully investigated in many of its theoretical and
observational aspects, the influence of a Cg component in some
kinematic tests such as the angular size-redshift relation
($\theta - z$) still remains to be studied. This is the goal of
the present paper. In the next sections we investigate qualitative and
quantitative aspects of the angular size-redshift test in
cosmological models driven by this dark matter/energy component. We first
investigate the influence of the Cg on the minimal redshift at
which the angular size of an extragalactic source takes its
minimal value. Afterwards, we consider the $\theta(z)$ data
recently updated and extended by Gurvits et al. (1999) to
constrain the equation of state of the Cg component as well as a
combination between these $\theta(z)$ observations and the latest
SNe Ia data, as provided by Riess {\it et al.} (2004). To do so,
we follow Alcaniz et al. (2002) and consider two different cases,
namely, a flat scenario in which the generalized Chaplygin gas
together with the observed baryonic content are responsible by the
dynamics of the present-day Universe [unifying dark matter-energy]
(UDME)  and a flat scenario driven by non-relativistic matter plus
the generalized Chaplygin gas (CgCDM). For UDME scenarios we adopt
in our computations $\Omega_b = 0.04$, in accordance with the
estimates of the baryon density at nucleosynthesis (Burles,
Nollett \& Turner 2001) and the latest measurements of the Hubble
parameter (Freedman et al. 2002). For CgCDM models we assume
$\Omega_{m} = 0.3$, as suggested by dynamical estimates on scales
up to about $2h^{-1}$ Mpc (Calberg et al. 1996; Dekel et al.
1997). For the sake of completeness an additional analysis for the
original Cg model ($\alpha = 1$) is also included.

This paper is organized as follows.  In Sec. II we present the basic equations and
distance formulas necessary for our analysis. Following the method developed in
Lima \& Alcaniz (2000a; 2000b) the influence of a Cg-like component on the minimal
redshift $z_m$ is investigated. In Sec. III we analyze the constraints on the equation
of state of the Cg component from measurements of the angular size of
compact radio sources and SNe Ia data and compare them with other independent limits. We
end the paper by summarizing the main results in the conclusion Section.

\section{Distance Formulas and the minimal redshift}

\subsection{Distance Formulas}

Let us now consider the flat FRW line element $(c=1)$
\begin{equation}
 ds^2 = dt^2 - R^{2}(t) [d\chi^{2} + \chi^{2} (d
 \theta^2 +
\rm{sin}^{2} \theta d \phi^{2})],
\end{equation}
where $\chi$, $\theta$, and $\phi$ are dimensionless comoving
coordinates. In this background, the angular size-redshift relation for a
rod of intrinsic length $\ell$ is easily obtained by integrating
the spatial part of the above expression for $\chi$ and $\phi$
fixed, i.e.,
\begin{equation}
\theta(z) = {\ell \over d_A}.
\end{equation}
The angular diameter distance $d_A$ is given by
\begin{equation}
d_A = \frac{R_o \chi}{(1 + z)} = {H_o^{-1}  \over (1 + z)} \int_{(1 + z)^{-1}}^{1}
{x^{-2} dx \over
E(\Omega_{j}, A_s, \alpha, x)}
 \quad  ,
\end{equation}
where $x = {R(t) \over R_o} = (1 + z)^{-1}$ is a convenient
integration variable. For the kind of models here considered, the
dimensionless function $E(\Omega_{j}, A_s, \alpha, x)$ takes the following form
\begin{equation}
E = \left\{\frac{\Omega_{j}}{x^{3}} + (1 - \Omega_{j})\left[A_s + \frac{(1 -
A_s)}{x^{3(\alpha + 1)}}\right]^{\frac{1}{\alpha + 1}}\right\}^{1/2},
\end{equation}
where $\Omega_{j}$ stands for the baryonic matter density
parameter ($j = b$) in UDME scenarios and the baryonic + dark
matter density parameter ($j = m$) in CgCDM models. Note that from
the above equation UDME models reduce to the $\Lambda$CDM case for
$\alpha = 0$ whereas CgCDM models get the same limit, regardless
of the value of $\alpha$, when $A_s = 1$. In both cases, the
standard Einstein-de Sitter behavior is fully recovered for $A_s =
0$ (Avelino et al. 2003b; Fabris et al. 2004).

\subsection{Minimal redshift}

As widely known, the existence of a minimal redshift $z_m$ on the
angular size-redshift relation may qualitatively be understood in
terms of an expanding space: the light observed today from a
source at high $z$ was emitted when the object was closer (for a
pedagogical review on this topic see Janis 1986). The relevant
aspect here is to demonstrate how this effect may be quantified in terms of the
Chaplygin parameters $\alpha$ and $A_s$. To analyze the sensivity
of the critical redshift to this dark component, we adopt here the
approach originally presented by (Lima \& Alcaniz 2000a; 2000b).
The numerical results of this method have been confirmed by Lewis
\& Ibata (2002) through a Monte Carlo analysis.

The redshift $z_{m}$ at which the angular size takes its
minimal value is the one cancelling out the derivative of $\theta$ with
respect to $z$. Hence, from Eqs. (4)-(6) we have the following condition
\begin{equation}
\chi_m = (1 + z_m)\chi'|_{z = z_m},
\end{equation}
where a prime denotes differentiation with
respect
to $z$ and by definition $\chi_{m}= \chi(z_{m})$.
Note that Eq. (5) can readily be differentiated, yielding
\begin{equation}
(1 + z_{m})\chi'|_{z = z_m}  =  \frac{1}{R_o H_o}  {\cal{F}}(\Omega_{j}, A_s, \alpha,
z_m),
\end{equation}
where
\begin{eqnarray}
{\cal{F}} & = & \{\frac{\Omega_{j}}{(1 + z_m)^{-3}} +
\nonumber \\
&    &
+ (1 - \Omega_{j})\left[A_s + \frac{(1 -
A_s)}{(1 + z_m)^{-3(\alpha + 1)}}\right]^{\frac{1}{\alpha + 1}}\}^{1/2}.
\end{eqnarray}
Finally, by combining equations (7)-(9), we find
\begin{equation}
\int_{(1 + z_m)^{-1}}^{1} {dx \over x^2 E(\Omega_{j}, A_s, \alpha, z_m)} =
{\cal{F}}(\Omega_{j}, A_s, \alpha, z_m).
\end{equation}

In Fig. 1 we show the results of the above expression. Panels (a) and (c) show the
minimal redshift
$z_m$ as a function of the parameter $A_s$ for selected values of $\alpha$ in the
context of
CgCDM ($\Omega_m = 0.3$) and UDME ($\Omega_b = 0.04$) models, respectively, whereas
Panels (b) and (d) display the
$\alpha - z_m$ plane for some values of $A_s$ also for CgCDM and UDME scenarios (note
that the
scale of $z_m$ in each panel is different). The
minimal redshift is a much more sensitive function to the parameter $A_s$ than to the
index $\alpha$.
As can be seen in the panels [and also expected from Eq. (5)], regardless
of the value of $\alpha$, models with $A_s = 0$ reduce to the Einstein-de Sitter case so
that the standard result $z_m = 1.25$ is fully recovered. As physically expected (see,
e.g., Lima \& Alcaniz 2000a; 2000b), the
smaller the contribution of the material component (baryonic and/or dark) the higher the
minimal redshift $z_m$. From this qualitative argument UDME models would be in a better
agreement with the observational data than CgCDM scenarios since the current data for
milliarcsecond radio sources do not show clear evidence for a minimal angular size
($\theta_{\rm{minimal}}$) close to $z = 1.25$ (Gurvits et al. 1999). For the best fit
CgCDM model obtained from an analysis involving galaxy clusters X-ray and supernova
data, i.e., $A_s = 0.98$ and $\alpha = 0.93$ we find $z_m \simeq 1.6$, a value that is
very similar to the one predicted by the current concordance model, namely, a flat
$\Lambda$CDM scenario with $\Omega_m = 0.3$ and also by the standard FRW model with
$\Omega_m = 0.5$ (see Table I of Lima \& Alcaniz 2000a). It is worth mentioning that if
high-redshift sources present cosmological evolution, such an effect would move the
position of the $\theta_{\rm{minimal}}$ to extremely high redshift .

\section{Constraints from angular size measurements}

In this section we study the constraints from angular size
measurements of high-$z$ radio sources on the free parameters of
the Cg model. In order to constrain the parameters $A_s$ and
$\alpha$ we use the angular size data for milliarcsecond radio
sources recently compiled by Gurvits {\it et al.} (1999). This
data set is composed by 145 sources at low and high redshifts
($0.011 \leq z \leq 4.72$) distributed into 12 bins with 12-13
sources per bin. Since the main difference between the analysis
performed in this Section and the previous ones that have used
these angular size data to constrain cosmological parameters is
the background cosmology, we refer the reader to previous works
for a more complete analysis and detailed formulas (see, for
instance, Jackson \& Dodgson 1997; 2003; Vishwakarma 2001; Lima \&
Alcaniz 2002; Zhu \& Fujimoto 2002; Jain et al. 2003; Chen \&
Ratra 2003).

Following a procedure similar to that described in the quoted reference, we determine
the cosmological parameters $A_s$ and $\alpha$ through a $\chi^{2}$ minimization
for a range of $A_s$ and $\alpha$ spanning the interval [0, 1] in steps of
0.02, i.e.,
\begin{equation}
\chi^{2}(l, \Omega_{j}, A_s, \alpha) =
\sum_{i=1}^{12}{\frac{\left[\theta(z_{i}, \ell, \Omega_{j}, A_s, \alpha) -
\theta_{obs_i}\right]^{2}}{\sigma_{i}^{2}}},
\end{equation}
where $\theta(z_{i}, l, \Omega_{j}, A_s, \alpha)$ is given by Eqs. (4)-(6) and
$\theta_{obs_i}$ is the observed values of the angular size with errors $\sigma_{i}$ of
the $i$th bin in the sample. 68$\%$ and 95$\%$ confidence regions are defined by the
conventional two-parameters $\chi^{2}$ levels 2.30 and 6.17, respectively.

Figures 2a and 2b show the binned data of the median angular size
plotted as a function of redshift for selected values of $A_s$ and
$\alpha$ for UDME and CgCDM models, respectively. For comparison
the current ``concordance" scenario, i.e, the flat $\Lambda$CDM
model with $\Omega_m = 0.3$ is also shown (thick line). In Figs.
3a and 3b, we show the 68\% and 95\% confidence contours computed
by fixing the characteristic length $D$ at its best fit value from
a minimization of Eq. (11) relative to the parameters $A_s$,
$\alpha$ and $D$. From this analysis we obtain the following best
fit values: $A_s = 0.96$, $\alpha = 0.84$ and $D = 24.1h^{-1}$ pc
($\chi^{2}_{min} = 4.44$) for UDME scenarios and $A_s = 1.0$ and
$D = 29.58h^{-1}$ pc ($\chi^{2}_{min} = 4.57$) for CgCDM models.
Both cases represent accelerating universes with deceleration
parameter and the total age of the Universe given by $q_o = -0.88$
and $t_o \simeq 11h^{-1}$ Gyr (UDME) and $q_o = -0.55$ and $t_o
\simeq 9.4h^{-1}$ Gyr (CgCDM).  Note that, as happens in the
calculation of the minimal redshift $z_m$ as well as in other
cosmological tests (see, e.g., Cunha, Lima \& Alcaniz
2004), the current angular size data constrain more strongly the
parameter $A_s$ than the index $\alpha$. From Fig. 3 it is
perceptible that while the parameter $A_s$ is constrained to be $>
0.68$ (UDME) and $> 0.54$ (CgCDM) at $2\sigma$ the entire interval
of $\alpha$ is allowed.

\subsection{$\theta(z)$ + SNe Ia}

By combining the angular and luminosity distances, interesting
constraints on the Cg parameters are obtained. To perform such
analysis, we follow the conventional magnitude-redshift test (see,
e.g, Goliath et al. 2001; Dicus \& Repko 2003; Alcaniz \& Pires 2004) and use the
latest SNe Ia data set that corresponds to the gold sample (157
events including 9 SNe at $z > 1$ ) of Riess {\it et al.} (2004).
In this analysis, both the characteristic angular size $D$ and the
Hubble parameter $H_o$ are considered ``nuisance" parameters so
that we marginalize over them. In the case of the $\theta(z)$ test
the marginalization over the characteristic length can be easily
done by defining a modified $\tilde{\chi}^2$ statistics as:
\begin{eqnarray}
    \tilde{\chi}^2&=&
    -2\ln\left[\int_{0}^\infty\,d\ell
    \exp\left(-\frac{1}{2}\chi^2\right)\right]\\
    &=& A-\frac{B^2}{C}+\ln\left(\frac{2C}{\pi}\right), \nonumber
\end{eqnarray}
where
\begin{equation}
A = {\theta_{obs}}^{2} \sum_{i=1}^n\frac{1}{\sigma_i^2},
\end{equation}
\begin{equation}
B = - \frac{\theta_{obs}}{d_A}\sum_{i=1}^n\frac{1}{\sigma_i^2}
\end{equation}
and
\begin{equation}
C = \frac{1}{d_A}\sum_{i=1}^n\frac{1}{\sigma_i^2}.
\end{equation}
For a similar procedure concerning the magnitude-redshift test we refer the reader to
(Goliath et al. 2001; Bertolami \& Silva 2003).

Figures (4) and (5) show the results of our analysis. In Panels 4a
and 4b we display contours of the combined likelihood analysis for
the parametric space $\alpha - A_s$ in the context of CgCDM and
UDME scenarios, respectively. For UDME models we see that the
available parameter space is considerably modified when compared
with Fig. 3b, with the best-fit value for $A_s$ provided by the
$\theta(z)$ analysis, i.e., $A_s = 0.96$ being off by $\sim
3\sigma$ relative to the joint analysis. This analysis also yields
$0.65 < A_s < 0.90$ ($95\%$ c.l.) for UDME models and $A_s > 0.85$
($95\%$ c.l.) for CgCDM scenarios. These particular limits on
$A_s$ are in good agreement with the ones obtained from quasar
lensing statistics ($A_s \geq 0.72$, Dev, Jain \& Alcaniz 2002),
the old SNe Ia data ($A_s = 0.87^{+0.13}_{-0.18}$, Fabris et al.
2002; Avelino et al. 2002) as well as from the location of the
acoustic peaks of CMB as given by BOOMERANG and Arqueops ($0.57
\leq A_s \leq 0.91$ for $\alpha \leq 1$, Bento et al. 2003).
However, they are only marginally compatible with the tight
constraint obtained from the expected number of lensed radio
source for the Cosmic All-Sky Survey (CLASS) statistical data
(Dev, Jain \& Alcaniz 2004). Other interesting limits from this
analysis are also obtained on the original version of Cg, i.e., by
fixing $\alpha = 1$. In this case, the plane $\Omega_{\rm{m}} -
A_s$ (Fig. 5a) is reasonably constrained with the best fit values
located at $A_s = 0.81$ and $\Omega_{\rm{m}} = 0.0$ with
$\chi^{2}_{min}/\nu \simeq 1.09$. This particular value of the
matter density parameter agrees with the one found by Fabris {\it
et al.} (2002) by using the old sample of SNe Ia from High-$z$
Supernova Project. In this Ref., the authors interpret such a
result as a possible backup to the idea of dark matter-energy
unification (UDME models). However, in the light of recent CMB
data, unification from the original Cg ($\alpha = 1$) seems to be
quite unrealistic since the location of the acoustic peaks as
given by WMAP and BOOMERANG is in conflict with the predictions of
this particular scenario (see, e.g., Carturan \& Finelli
2003). Figure 5b shows the same analysis of Panel a by assuming
the Gaussian prior on the matter density parameter, i.e.,
$\Omega_m = 0.27 \pm 0.04$ (Spergel et al. 2003). As can be seen,
the parameter space is tightly constrained with the best-fit
scenario located at $\Omega_m = 0.26$, $A_s = 0.97$ and
$\chi^{2}_{min}/\nu \simeq 1.1$. In Table I we present the best
fit values for the Cg parameters obtained from the two main
analyses performed in this paper.

\section{Conclusion}

Based on a large body of observational evidence, a consensus is
beginning to emerge that we live in a flat, accelerated universe
composed of $\sim$ 1/3 of matter (barionic + dark) and $\sim$ 2/3
of a negative-pressure dark component. However, since the nature
of these dark components (matter and energy) is not well
understood, an important task nowadays in cosmology is to
investigate the existing possibilities in the light of the current
observational data. In this paper we have focused our attention on
some observational aspects of cosmologies driven by an exotic dark
energy component named generalized Chaplygin gas. These models
also constitute an interesting possibility of unification for dark
matter/energy (where these two dark components are seen as
different manifestations of a single fluid). Initially, We have
investigated the influence of the Cg on the minimal redshift
($z_m$) at which the angular sizes of extragalactic sources takes
their minimal values. From this analysis it was showed that the
location of the minimal redshift is a much more sensitive function
to the parameter $A_s$ than to the index $\alpha$.
By using a large sample of milliarcsecond radio sources recently updated and
extended by Gurvits {\it et al.} (1999) along with 
the latest SNe Ia data, as given by Riess {\it et al.} (2004) we obtained, as the best
fit for these data, $A_s =
0.84$ and $\alpha = 1.0$ (UDME) and $A_s = 0.99$ and $\alpha =
1.0$ (CgCDM). Such values are fully disagreement with the CMB analysis by Amendola {\it
et al.} (2004) which showed that Cg scenarios with $\alpha > 0.2$ are ruled out by the
current WMAP data. Finally, it should be remarked that the background results
presented here may be somewhat modified if one takes into account
the onset of the non-linear regime, as recently discussed by Avelino {\it et al.}
(2004).

\acknowledgments

The authors are very grateful to L. I. Gurvits for sending his compilation of the data.
This work is supported by the Conselho Nacional de Desenvolvimento Cient\'{\i}fico e
Tecnol\'{o}gico (CNPq - Brasil) and CNPq (62.0053/01-1-PADCT III/Milenio).

\clearpage

\begin{figure}
\epsscale{.80}
\rotatebox{-90}{\plotone{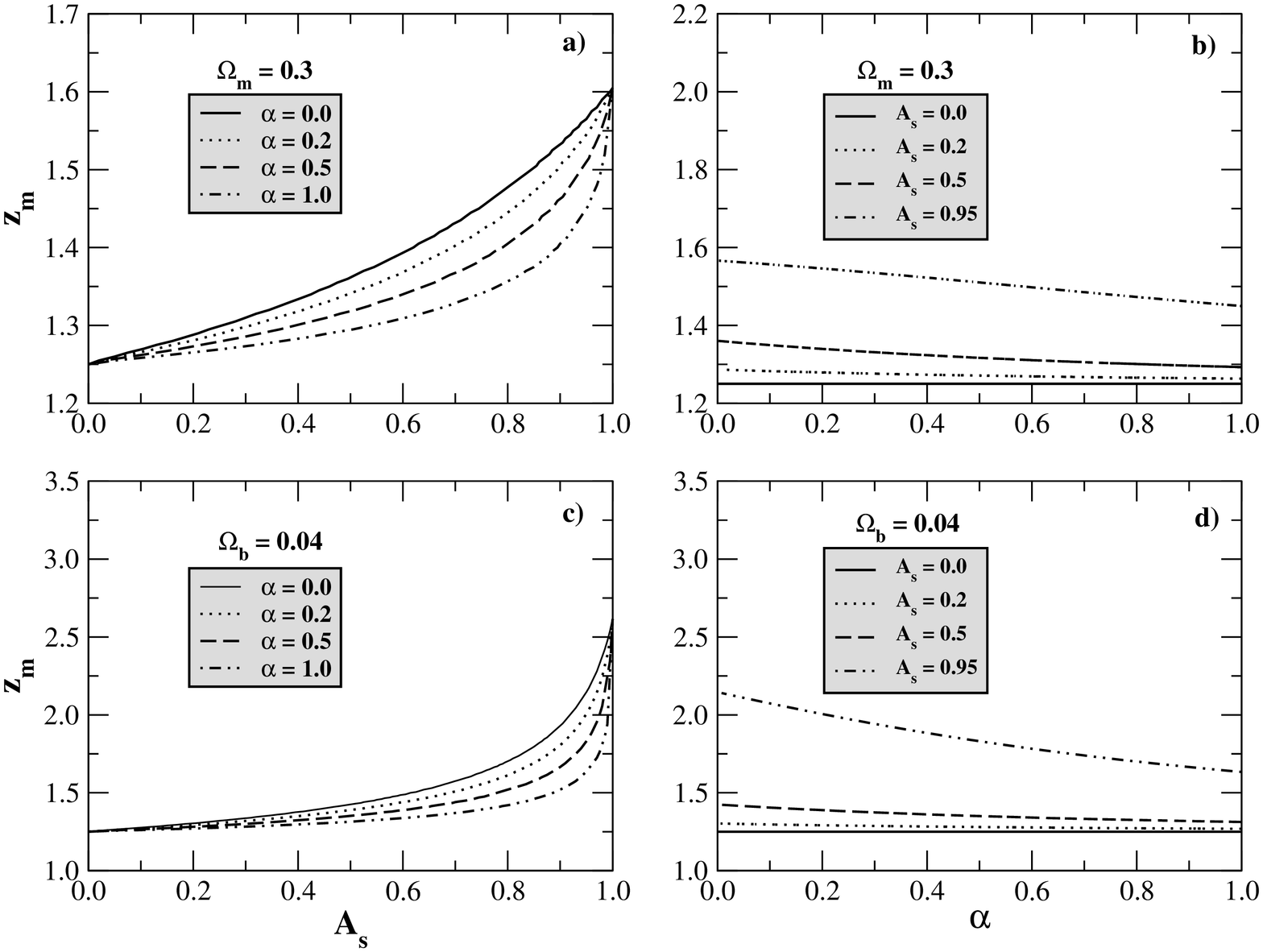}}
\caption{Critical redshift $z_m$ as a function of the Cg parameters $A_s$ and $\alpha$
for CgCDM (Panels a and b) and UDME (Panels c and d) scenarios. Note that, regardless
of the value of $\alpha$, all models with $A_s = 0$ recover the standard Einstein-de
Sitter result, i.e., $z_m = 1.25$.}
\end{figure}

\clearpage

\begin{figure}
\epsscale{.80}
\rotatebox{-90}{\plotone{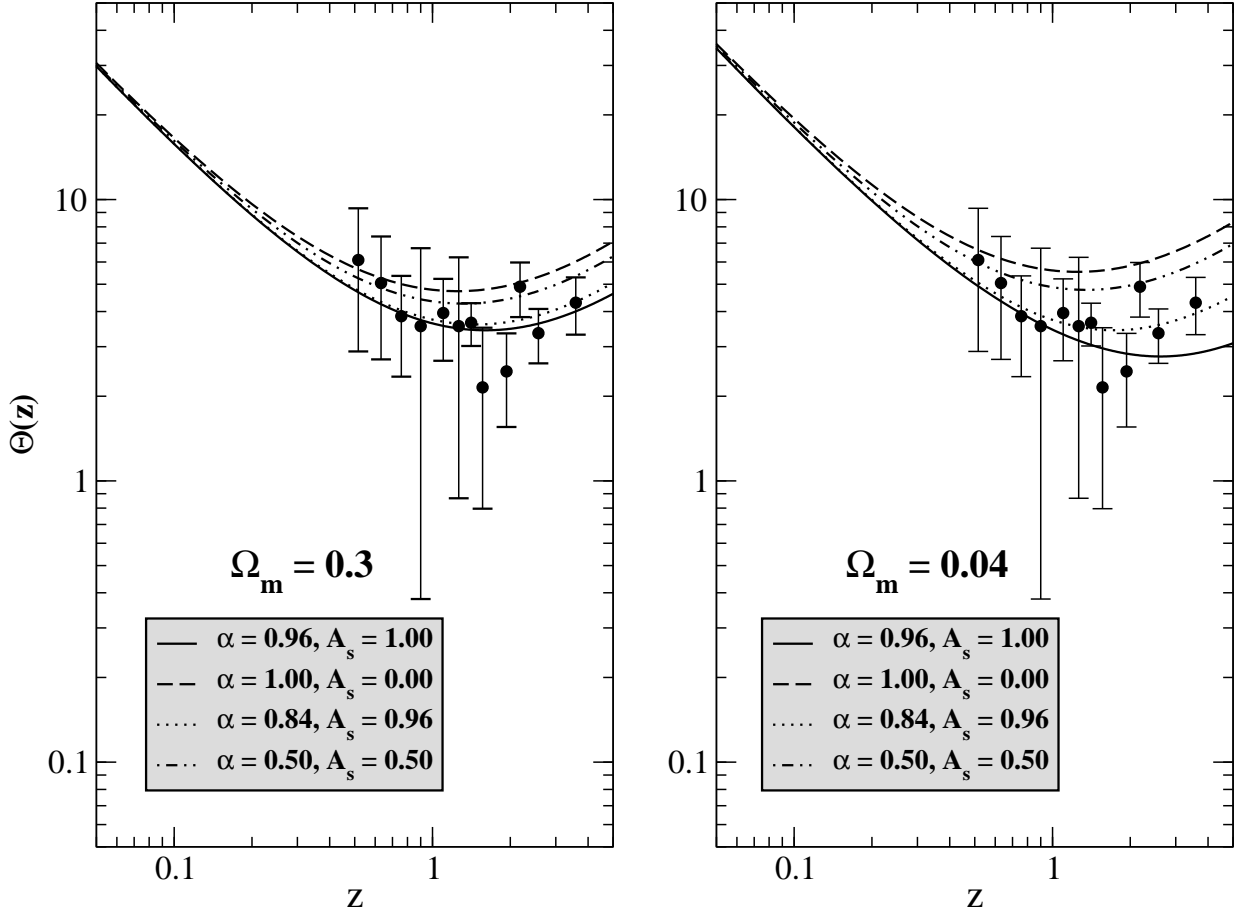}}
\caption{Angular size versus redshift for some
combinations of the parameters $A_s$ and $\alpha$ for CgCDM and
UDME models. The characteristic length $D$ has been fixed at
$29.58h^{-1}$ pc (Panel a) and at $24.10h^{-1}$ pc (Panel b),
which corresponds to the best fit values obtained from a
minimization of Eq. (11) relative to the parameters $A_s$,
$\alpha$ and $D$.}
\end{figure}

\clearpage

\begin{figure}
\epsscale{.80}
\plotone{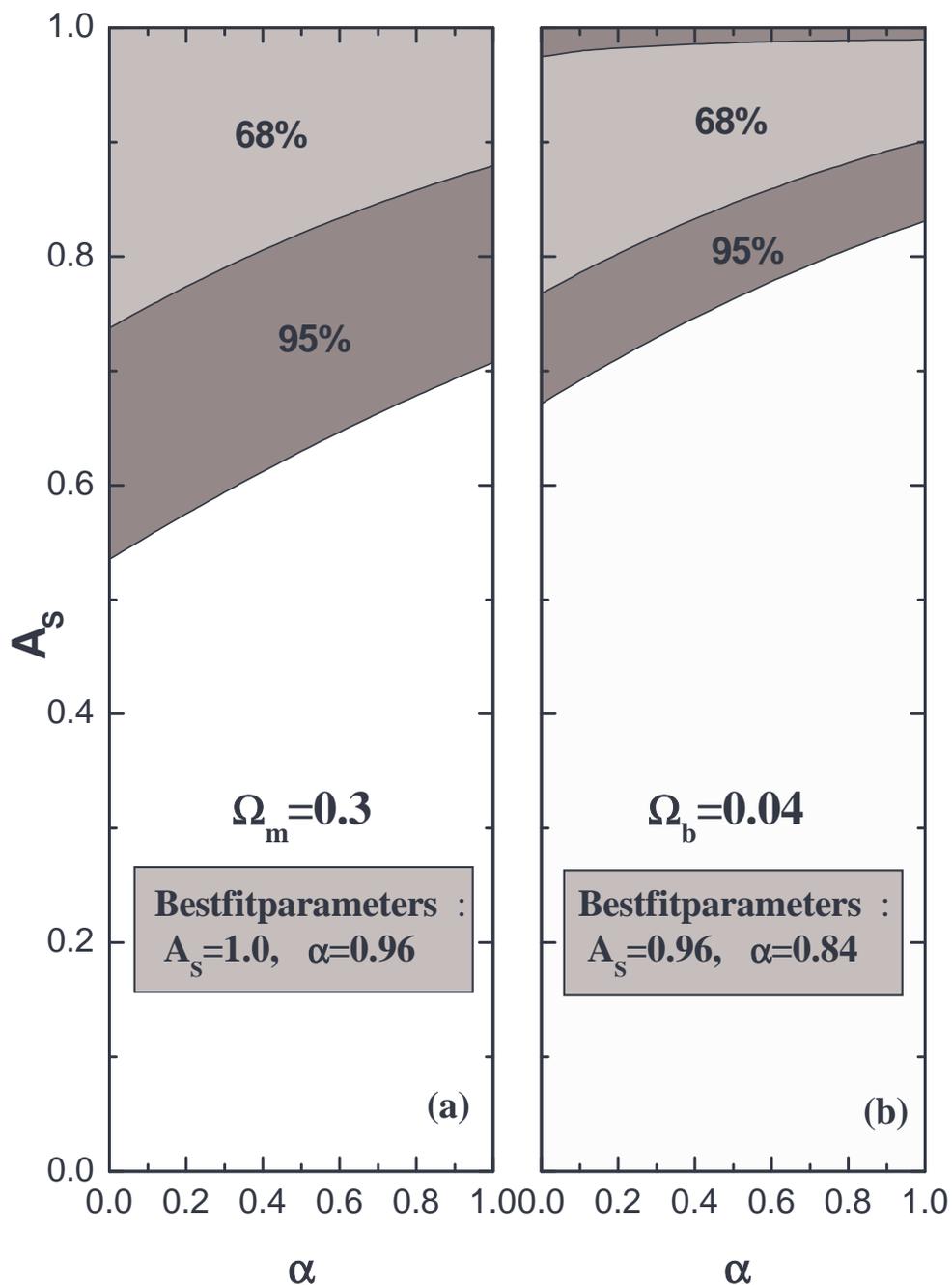}
\caption{Confidence regions in the $\alpha - A_s$ plane according to the
updated sample of angular size data of Gurvits {\it et al.} (1999) for CgCDM (Panel
a) and UDME (Panel b) models. From this figure we see that while the entire interval
of $\alpha$ is allowed the parameter $A_s$ is constrained to be $> 0.54$ (CgCDM) and $>
0.68$ (UDME) at $95\%$ confidence level.}
\end{figure}

\clearpage

\begin{figure}
\epsscale{.80}
\plotone{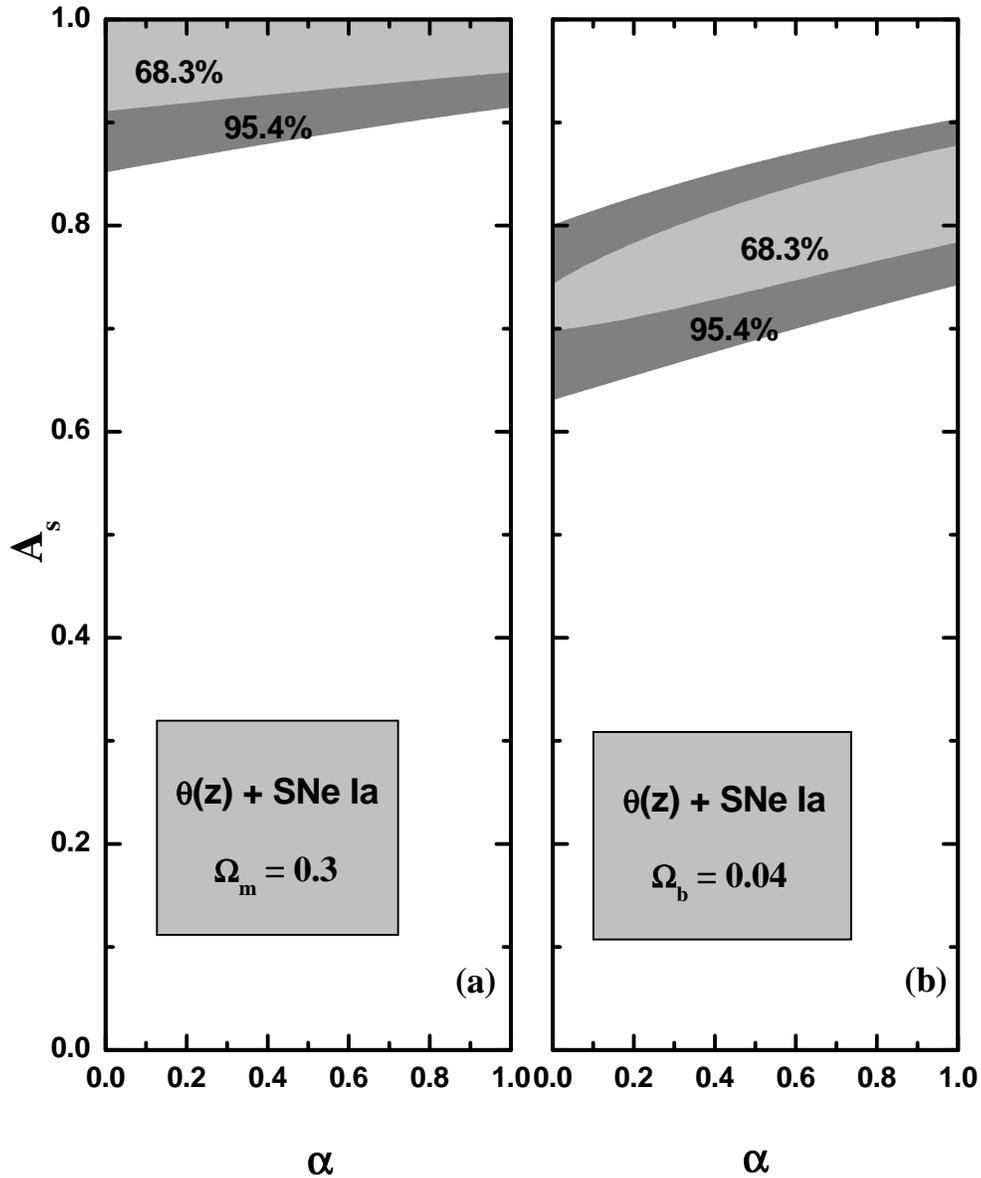}
\caption{{\bf{a)}} 68\% and 95\% confidence levels in the plane $\alpha - A_s$ for CgCDM
models corresponding to the joint $\theta(z)$ + SNe analysis as described in the text.
As in Fig. 3, the entire interval of $\alpha$ is allowed. The parameter $A_s$ is now
constrained to be $> 0.85$ at $95\%$ c.l. {\bf{b)}} The same as in Panel a for
UDME scenarios. For this case we find $0.65 < A_s < 0.90$ ($95\%$ c.l.)}
\end{figure}

\clearpage

\begin{figure}
\epsscale{.80}
\plotone{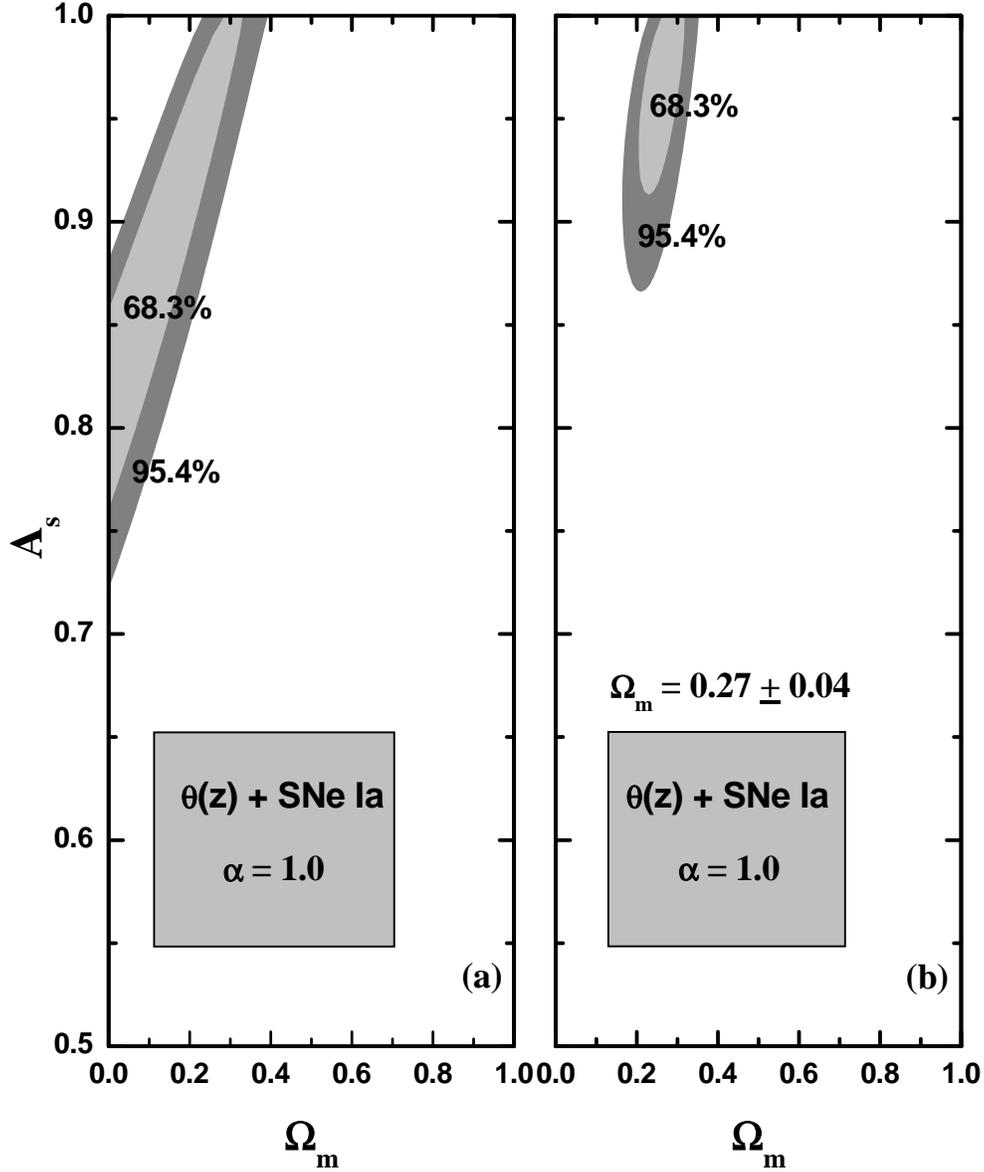}
\caption{{\bf{a)}} The $\Omega_m - A_s$
plane for the joint $\theta(z)$ + SNe analysis having $\alpha = 1$ (original version of
Cg). Here the best fit values are located at $A_s = 0.81$ and $\Omega_{\rm{m}} = 0.0$.
{\bf{b)}} The same as in Panel a by assuming a Gaussian prior on the matter density
parameter $\Omega_m = 0.27 \pm 0.04$ (Spergel et al. 2003). The best-fit scenario is
located at $\Omega_m = 0.26$ and $A_s = 0.97$ with $\chi^{2}_{min}/\nu \simeq 1.1$.}
\end{figure}
\clearpage
\clearpage

\begin{table}[h]
\caption{Best fit values for Cg parameters}
\begin{tabular}{lcl}
Test& UDME & $\quad      \quad     \quad  $CgCDM\\ \hline \hline
\\ $\theta(z)$.................& $A_s = 0.96 \quad  \alpha = 0.84$
& $A_s = 1.0 \quad \alpha = 0.96$\\ SNe.................& $A_s =
0.84 \quad  \alpha = 1.0$ & $A_s = 0.98 \quad \alpha = 0.96$\\
$\theta(z)$ + SNe.....& $A_s = 0.84 \quad  \alpha = 1.0$  & $A_s =
0.99 \quad  \alpha = 1.0$\\ \hline \hline \\
\end{tabular}
\end{table}

\end{document}